\documentstyle[prl,aps,epsfig,preprint]{revtex}

\draft

\newlength{\textwidthm}
\begin{document}
\title{Entanglement of Atomic Ensembles by Trapping Correlated 
Photon States}
\author{M.~D.~Lukin$^1$, S.~F.~Yelin$^1$ and M.~Fleischhauer$^2$}
\address{$^1$ ITAMP, Harvard-Smithsonian Center for Astrophysics,
                Cambridge, MA~~02138 }
\address{$^2$ Sektion Physik, Ludwig-Maximilians-Universit\"at M\"unchen,
Theresienstr. 37, D-80333 M\"unchen, Germany}
\date{\today}
\maketitle
\begin{abstract}
We describe a general
technique that allows for an ideal transfer of 
quantum correlations between light fields and metastable states of 
matter.  The technique is based on trapping quantum states of
photons in coherently driven atomic media, in 
which the group velocity is adiabatically reduced to zero. 
We discuss possible applications such as quantum state memories, 
generation of squeezed atomic states, preparation of 
entangled atomic ensembles and quantum information processing. 
\end{abstract}


One of the most intriguing aspects of quantum theory is the 
possibility to entangle quantum states of distinct objects.
Recently these ideas led to many interesting
new concepts such as quantum cryptography \cite{cryptography},
teleportation \cite{teleportation} and quantum computation
\cite{Q-comp}. Photons are the fastest, simplest and very robust carriers
of quantum information\cite{Gisin_exp}, but they are difficult to 
store.  

This Letter describes a general
technique that 
allows to transfer quantum correlations from 
traveling-wave light fields to collective atomic states and vice 
versa with nearly ideal efficiency. This is achieved by adiabatically
reducing the group velocity of light fields to zero, thereby ``trapping''
the photons in a medium. Specifically, we here  use the  technique 
of intracavity Electromagnetically Induced Transparency (EIT) 
\cite{ol98,dark}, 
in which the  properties of a cavity filled with three-state 
$\Lambda$-type atoms can be manipulated by an external (classical) field.  

Once the transfer is completed, the atomic ensemble ``stores'' all 
photons including their quantum correlations in metastable
superpositions of many-atom states. Consequently a procedure 
of this kind can be used to generate non-classical states of the 
atoms, and, in particular, to entangle two or more distant 
atomic ensembles by mapping entangled photon wave-packets onto 
spatially separated atomic systems. 
In addition to fundamental aspects, this makes applications
in  low-noise spectroscopy \cite{spe} and quantum 
teleportation of collective atomic states \cite{teleportation} feasible. 
Furthermore, the atomic excitations 
can be coherently manipulated over a very long period of time, 
which opens up very interesting possibilities 
for quantum information processing \cite{Q-comp}.  Finally, the
stored quantum states can be transfered back to light beams 
by simply reversing the adiabatic storage procedure. 

The present contribution is motivated by recent experiments, in which 
EIT  has been used to dramatically reduce the group velocity of 
light pulses in a coherently driven, optically dense ensemble of atoms 
\cite{group}. Ideas involving adiabatic passage  have already been 
considered for manipulation of single atoms in the  
context of cavity QED \cite{QED}. The method described here involves 
an optically dense many-atom  system and
does  not require the usual strong-coupling regime of cavity QED. 
In the present system single photons couple 
to {\it collective excitations} associated with EIT, and the corresponding 
coupling strength can exceed that of an individual atom by orders of 
magnitude. In contrast to the approaches  
involving partial transfer of correlations by dissipative 
absorption of light \cite{Polzik},  the present method is completely 
coherent and reversible.  

To illustrate the essence of the technique, consider a single-mode
cavity filled with a large number of
coherently driven $\Lambda$-type atoms as shown in Fig.~1a. One transition 
is coupled by the quantum cavity field, whereas
the other is driven by a classical coherent field of 
Rabi-frequency $\Omega(t)$. Under the condition of two-photon resonance 
the coherent driving field creates induced transparency \cite{dark}
for the cavity field and
the associated linear dispersion can substantially
reduce its group velocity \cite{group}.
This leads to a dramatic enhancement of the effective
storage time limited only by the lifetime of the 
dark state \cite{ol98}. The basic Hamiltonian of the cavity + $N$-atom 
system can be written in terms of collective 
atomic operators 
$\hat\Sigma_{ab} = \sum_{i = 1}^N  \hat\sigma_{ab}^i$ and $\hat\Sigma_{ac} = 
\sum_{i = 1}^N  \hat\sigma_{ac}^i$ as 
\begin{equation}
H = \hbar g \hat a\hat\Sigma_{ab} + 
\hbar\Omega(t)   
\hat\Sigma_{ac} + {\rm h.c.} ,   
\label{ham}
\end{equation}
where $\hat\sigma_{\mu\nu}^i = |\mu\rangle_{ii}\langle \nu|$ is the 
flip operator of the $i$th atom between states $|\mu\rangle$ and $|\nu\rangle$.
$g$ is the coupling  constant between the atoms and the field mode 
(vacuum Rabi-frequency) which for simplicity is assumed 
to be equal for all atoms. Here and below we work in a frame rotating with 
optical frequencies. 
This Hamiltonian has a family of dark 
states that are decoupled from both optical fields
\begin{equation}
|D,n\rangle = \sum_{k=0}^n \sqrt{{n!\over k! (n-k)!}} {
(-g)^k N^{k/2}\Omega^{n-k} 
 \over (g^2 N +\Omega^2)^{n/2}} |n-k\rangle|c^k\rangle
\label{dark}
\end{equation}
composed of cavity field states containing $|n-k\rangle$ photons and 
symmetric Dicke-like atomic states $|c^k\rangle$ containing $k$ atoms in 
level $|c\rangle$, and all others in the ground state $|b\rangle$:
\begin{eqnarray} 
|c^0\rangle &\equiv& |b_1...b_N\rangle, \quad
|c^1\rangle \equiv \sum_{i = 1}^N 
\frac{-1}{\sqrt{N}}\,
|b_1...c_i...b_N\rangle, \\
|c^2\rangle &\equiv& \sum_{i\ne j = 1}^N 
\frac{1}{\sqrt{2N(N-1)}}\, |b_1...c_i...c_j....b_N\rangle,  \; {\rm etc}.
\end{eqnarray} 
We here assumed that the number
of atoms is much larger than the number of photons in the 
light field.  

The essence of the adiabatic transfer is the asymptotic behavior of the 
dark states (\ref{dark}) in the two limiting cases: 
\begin{eqnarray} 
|D,n\rangle \rightarrow |n\rangle |c^0\rangle,\; {\rm when} \; \Omega \gg
g\sqrt{N}, \\
|D,n\rangle \rightarrow |0\rangle |c^n\rangle,\; {\rm when} \; \Omega \ll
g\sqrt{N}.  
\end{eqnarray} 
For a sufficiently strong coherent driving field the
atoms do not interact with light, and the dark state coincides
with the ``bare'' cavity mode where all atoms remain in the ground state.
In this limit photons can ``leak'' in and out of the cavity as if 
it would be empty. In the opposite limit, the
dark state is a purely atomic (metastable) state with no photons in
the cavity. In the latter case the lifetime of excitations will
not be sensitive to  
cavity decay; it will be limited solely by the decay of the metastable atomic
 states. It is most important that by varying the strength of the
driving field $\Omega(t)$, and consequently by changing the linear
dispersion in the  
medium, the state of the combined atom+cavity system can be changed from
cavity-like (in which excitation is mostly of photon nature) to the
atom-like (in which excitations are shared among the atoms). Since all dark 
states are orthogonal to each other, the 
ideal storing procedure (as discussed below) will transform any 
superposition of
photon states into corresponding superpositions of atomic states:
\begin{equation}
\sum_i \alpha_i\, 
|i\rangle|c^0\rangle\rightarrow \sum_i \alpha_i\, |0\rangle|c^i\rangle.
\label{tr1}
\end{equation}

Before proceeding with a detailed description of the specific adiabatic 
technique we note that the above results can be easily generalized to the 
case of two  atomic ensembles, which can be entangled by trapping two
entangled photon fields. In this case the atomic
ensembles are placed either within the same or 
two different optical cavities (Fig.1b). The dark states are then the direct
product of those corresponding to the subsystems:
$|D,n,m\rangle = |D_r,n_r\rangle |D_l,m_l\rangle.$   
Hence the following operation can be accomplished:
\begin{equation}
\sum_{nm} \alpha_{nm}\, |nm\rangle|c_r^0 \rangle 
|c_l^0\rangle\rightarrow \sum_{nm} 
\alpha_{nm} |0\rangle|c_r^n\rangle |c_l^m\rangle.
\label{tr2}
\end{equation}
It is clear that trapping perfectly entangled photon states will 
result in perfectly entangled atomic ensembles. 

Yet another related situation that we wish to mention involves trapping 
the states of two distinct fields within {\it one} atomic species. Here the 
two fields interact with more complex atoms, such as 
those shown in Fig.1c. By using essentially the same arguments 
as above one finds that a perfect state-transfer of the  two fields 
to the atoms yields: 
\begin{equation}
\sum_{nm}\alpha_{nm}\, |nm\rangle|c_r^0 c_l^0\rangle\rightarrow \sum_{nm} 
\alpha_{nm}\, |0\rangle|c_r^nc_l^m\rangle.
\label{tr3}
\end{equation}
The potential significance of the last scheme is that the 
correlations and the entanglement 
of the two fields can be manipulated, since they are now stored within the same
atomic ensemble. 
This is of prime importance in quantum information processing,
in particular, for quantum logic devices and quantum computation.

We now describe and analyze an adiabatic procedure by which  an 
input traveling-wave quantum field can be captured, stored and released. 
To this end we consider a quasi-1D system,
include the continuum of the free-space plane-wave modes (with creation 
operators $b^+_k$) and model the coupling of these modes with the cavity by
an effective Hamiltonian $ V = 
\hbar\sum_k \kappa {\hat a}^\dagger {\hat b}_k + {\rm h.c.}$.  $\kappa$
is a coupling constant. The initial
state of the free field is taken to be $ |\Psi_{\rm in}\rangle = 
\sum_k\xi_k^1|1_k\rangle + 
\sum_{k,m} \xi_{k,m}^2|1_k1_m\rangle + ...\;$. 
It is convenient to work with correlation amplitudes, i.e. 
Fourier transforms of $\xi_{k...l}^j$:
\begin{eqnarray}
\Phi_j(t_1...t_j) = \langle 0| {\hat E}(t_1)...{\hat E}(t_j) |\Psi\rangle,  
\end{eqnarray}
where ${\hat E}(t) = L/(2\pi c) \int d\omega_k \exp(i\omega_kt) {\hat b}_k$,
and $L$ is quantization length. 
E.g. $\Phi_1$ describes the envelope of a single photon wave packet, 
$\Phi_2$ is the coincidence amplitude etc.  
We now consider a broad class of pulsed fields that are characterized
by a single envelope $h(t)$ such that 
\begin{eqnarray}
\Phi_j(t_1,t_2,...t_j) = \alpha_j \sqrt{j!} h(t_1) h(t_2)...h(t_j).
\end{eqnarray} 
A quantum state of such pulses can be described by a density matrix
$\rho_{nm}=\alpha_n^*\alpha_m$. 
The corresponding 
mode function (envelope) is a superposition of plane waves proportional to
$h(z/c)=\int{\rm d}\omega_k\, \xi_k\, {\rm e}^{i\omega_k z/c}$.
When the pulses interact with the 
combined system of cavity mode and atoms the states $\alpha_j|c^j\rangle$
are excited. 
We proceed by deriving the equations of motion for 
the probability amplitudes in
the basis of dark and orthogonal bright states. The 
bright states as well as the excited states (containing components of 
states $|a^i..\rangle$) are then adiabatically eliminated. 
The remaining amplitudes 
of dark states and free-field components form a Dicke-like ladder. The 
ladder states are coupled to each other with the time-dependent
coupling strength $\kappa \cos\theta(t)$ determined by the mixing angle 
$\cos\theta(t) = \Omega(t)/ \sqrt{\Omega(t)^2 + g^2 N}$.

In the case when only single-photon pulses are involved the evolution 
equations are \cite{Marlan_fest}: 
\begin{eqnarray}
{\dot{  D}_1}(t) &=& i \kappa\cos\theta(t)\sum_k{ \xi}_k(t),
\label{D1}\\
{\dot{  \xi}}_k(t) &=& -i\Delta_k\, { \xi}_k(t) + i \kappa
\cos\theta(t)\, {  D}(t).\label{xi}
\end{eqnarray}
We proceed by formally integrating Eq.(\ref{xi}), substituting the 
result into Eq.(\ref{D1}) and invoking the standard Markov approximation. 
Assuming that no photons arrive to the cavity before $t_0$ we
find for the dark state amplitude $D_1(t) = - i \alpha_1 d(t)$ with 
\begin{eqnarray}
{  d}(t)&=&\sqrt{\gamma {c \over L}}\int_{t_0}^t\!\!{\rm d}\tau\,
\cos\theta(\tau)\, h(\tau) \nonumber \\
&\times& \exp\left\{-\frac{\gamma}{2}\int_\tau^t\!\!{\rm d}\tau^\prime
\, \cos^2\theta(\tau^\prime)\right\},\label{Dres}
\end{eqnarray}
and for the input-output relation   
\begin{equation}
h_{\rm out}(t) = h(t) - \sqrt{\gamma L/c}\,  d(t), 
\label{out}
\end{equation}
where $h_{\rm out}(t)$ is a pulse-shape of the outgoing wave packet. 
Here we have introduced the empty-cavity decay rate $\gamma=\kappa^2 L/c$.
In order to trap photons  we require 
$h_{\rm out}(t) = {\dot h}_{\rm out}(t)=0$. 
Differentiating 
Eqs.(\ref{Dres},\ref{out})  yields
\begin{eqnarray} 
-\frac{{\rm d}}{{\rm d} t}  \, {\rm ln}\, \cos\theta(t)
+\frac{{\rm d}}{{\rm d} t} \, {\rm ln}\, h(t)
=\frac{\gamma}{2}\cos^2\theta(t)
,\label{impedance}
\end{eqnarray}
and with the asymptotic condition $\cos\theta \to 0$ 
the output field remains zero. 
The above condition corresponds to a quantum or dynamical impedance matching 
\cite{Marlan_fest}.
The term on the r.h.s. is the effective
cavity decay rate reduced due to intracavity EIT \cite{ol98}.
The first term on the l.h.s. of Eq.(\ref{impedance}) describes 
internal ``losses'' due to  coherent Raman
adiabatic passage 
and the second term  is due to the time-dependence of the input field.
As in the case of classical impedance matching \cite{Siegman}, 
Eq.~(\ref{impedance}) reflects the condition for
complete destructive interference resulting in a vanishing outgoing wave. 
 Solving Eq.(\ref{impedance}) yields
\begin{equation}
\cos^2\theta(t) =\displaystyle{\frac{h^2(t)}{
\gamma\int_{-\infty}^t\!{\rm d}\tau h^2(\tau)}},
\label{imp_sol}
\end{equation}
which corresponds to $d(t\rightarrow +\infty) \rightarrow 1$ (see Fig.2a).
Hence, by suitable variation of the classical driving field 
any single-photon pulse can be trapped ideally,
if its pulse length is longer than the bare-cavity decay time.

Generalizations of the above considerations to multi-photon states can proceed
along the same lines, but involve more tedious algebra. In particular, 
for the two-photon states one finds  $D_2(t) = - \alpha_2 d(t)^2$, and in general
\begin{equation}
D_k(t) = (-i)^k \alpha_k d(t)^k
\end{equation}
can be proved. Under conditions of quantum impedance
matching $D_k(t\rightarrow \infty)\rightarrow (-i)^k a_k$ for arbitrary $k$. 
Hence pulsed fields in a generalized single mode 
with arbitrary quantum state can be mapped onto
the atomic ensemble.

Releasing the stored quantum state into a pulse of desired shape can be 
accomplished in a straightforward way. A simple reversal of the time 
dependence of the control field at a later time $t_d$ leads to a
perfect mirror-image of the 
initial pulse. This can be verified directly from Eqs. 
(\ref{Dres},\ref{out}), see also Fig.2a. However, $\cos\theta$ can also be
rotated back to its original value in another way, which allows to
``tailor'' the pulse-shape of the outgoing wave-packet while
retaining the quantum state.     

We next examine the factors limiting the performance of the photon 
trapping scheme. Adiabatic following occurs if the population in the 
excited and bright states is small at all times. 
The corresponding  
conditions can be derived by substituting the adiabatic solutions into 
the exact equations and requiring that the coupling of dark states to all 
other states is small \cite{adiab}.   For a pulse
duration $T$, a linewidth of the excited state $\gamma_a$ and a cavity width 
$\gamma$, the adiabaticity conditions are 
\begin{equation}
\Omega(t)^2+ g^2 N \gg {\rm max}\; 
\biggl[\gamma\gamma_a,\;\frac{\gamma_a}{T},\; 
\sqrt{\frac{\gamma}{T}}\, \gamma_a\biggr].
\label{adiabat_cond}
\end{equation}
Since the characteristic input-pulse length and thus the characteristic
times $T$ have to be larger or equal to the bare cavity decay time 
$\gamma^{-1}$, the first condition is the most stringent one. Therefore 
adiabatic following is possible provided that $g^2 N \gg \gamma\gamma_a$.

In the discussion above we have disregarded the finite lifetime of the
metastable state, $\gamma_0^{-1}$.  
If $\gamma_0$ is small, its influence during the loading and 
unloading periods can be neglected but needs to be taken into account
 during the storage interval. Collective states such as $|c^n\rangle$
will dephase at a rate $\gamma_n = n \gamma_0$, which sets the upper 
limit on the longest storage time. 
To illustrate the effect of this damping we have plotted in Fig.~2b
the fidelity 
of the quantum state storage, 
defined as $f={\rm Tr}\, \{\rho_{\rm in}\rho_{\rm out}\}$,
as function of the storage time $t_s$ (the time between capture and release) 
for input pulses in a number state and a squeezed vacuum state.
It is apparent that the maximum storage time is on the order of the
single-atom decay time divided by the characteristic
number of input photons. We note that in alkali-vapor cells with
buffer gas and/or wall coatings dark state lifetimes on the order of 
seconds are observed e.g. by Budker et al.\cite{group}.

We conclude by summarizing the main results and outlining the 
possible avenues of future studies opened by the present work. 
We demonstrated that is possible to map ideally the 
quantum states of light fields onto metastable states of atomic ensemble. In 
particular, this allows for the generation of non-classical (e.g. squeezed)
states of atoms (see Eq.\ref{tr1}). These states are precisely of the form 
required to achieve spectroscopic sensitivity beyond the usual quantum 
limit \cite{spe}. We have shown that by trapping two correlated fields 
it is possible to generate entangled states of two distinct atomic ensembles 
(see Eq.\ref{tr2}). Prospects for quantum teleportation of collective atomic 
states are hence feasible. Finally, by trapping two fields within the same 
multi-state species (see Eq. \ref{tr3}) it seems possible to create good
conditions for coherent manipulation of field correlations, entanglement, 
etc. This opens up interesting prospectives for quantum information 
processing and, in  particular, for quantum logic operations. 
It is essential that 
all of the transfer operations can be achieved without invoking 
the strong coupling regime of single-atom cavity QED. Specifically, we 
have shown that under conditions of quantum impedance matching free
fields can be ideally transfered back and forth provided the
excitation rate of  
the collective mode ($\sim g^2N/\gamma_a$) exceeds the cavity decay rate 
(see Eq.\ref{adiabat_cond}).  Potentially, this can be used to considerably 
improve  the fidelity of quantum processing. 

At the same time we note that several interesting questions remain open
and need to be explored. For example, we have not considered here any specific 
schemes to perform quantum logic gates with trapped photons. Possible 
ways include cavity QED techniques \cite{QED}, direct nonlinear 
interactions of photons
via e.g. resonantly enhanced Kerr nonlinearities \cite{NLO} or, alternatively, 
atom-atom  interactions. Another way 
of manipulating the quantum statistics involves the regime in which, contrary 
to the situation considered here, the characteristic 
number of photons exceeds the number of atoms.
We further note 
that although the present analysis involves electronic degrees of freedom, 
it can be used to excite states of the 
center-of-mass motion of cold atomic samples in BECs \cite{Parkins99}. 
Here again
atom-atom collisions need to be taken into account. This adds
new interesting dimensions to the present studies and will discussed 
elsewhere.

We thank A.Imamo\u glu, R. Glauber and W.~Vogel for many stimulating 
discussions. This work was supported by the National Science Foundation. 
S.F.Y. thanks Humboldt Foundation for support.

\def\etal{\textit{et al.}}


\begin{figure}[ht]
\centerline{\epsfig{file=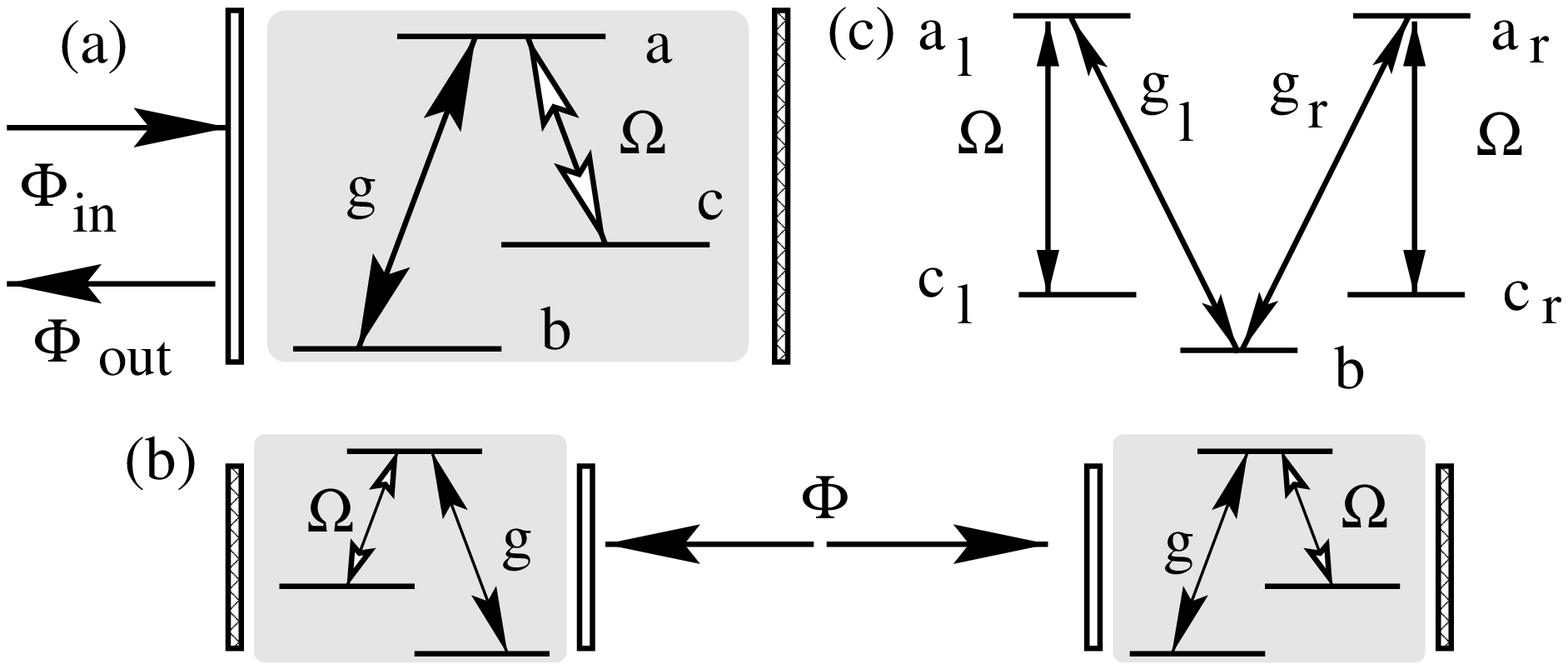,width=8.6cm}}
 \vspace*{2ex}
 \caption{ (a) Optical cavity with single out-coupling
mirror filled with large number of $\Lambda$-type atoms. External
coherent field of Rabi-frequency $\Omega(t)$ is used to dynamically
control properties of resonator system. Input field state is transferred
back and forth to atomic system via  coherent Raman adiabatic passage from
states $|b\rangle$ to state $|c\rangle$. (b) Generation 
of entangled atomic ensembles in distant cavities using correlated
 photons. (c) Multi-state atoms for trapping 
correlated photons of right and left circular polarizations. }
\end{figure}


\begin{figure}[ht]
\centerline{\epsfig{file=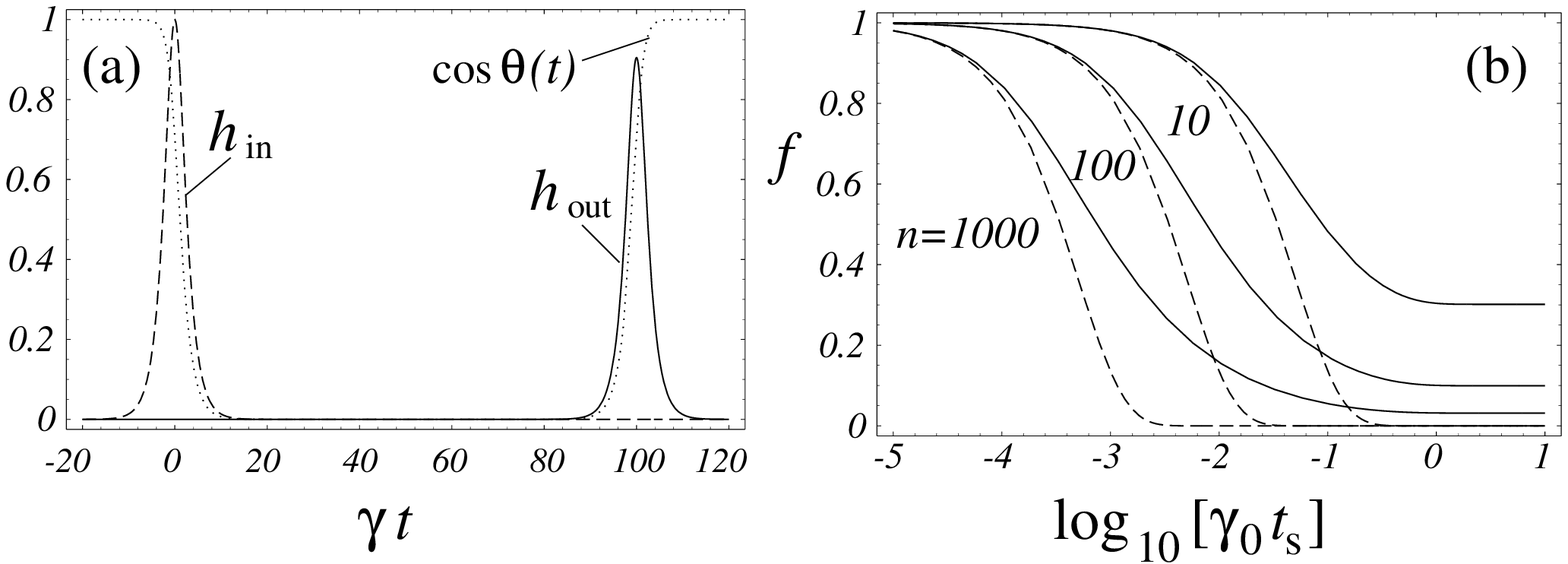,width=8.6cm}}
 \vspace*{2ex}
 \caption{(a) Storage of a hyperbolic secant pulse.
Shown are normalized input (dashed line) and output pulse (full line)
 as a function of time as well as
time dependence of $\cos\theta(t)$, optimized
for the input field. Time unit is decay time of bare-cavity $\gamma^{-1}$.
Dark-state decay rate is $\gamma_0=10^{-3}\gamma$.
(b) Fidelity of storage
for Fock state (dashed line) and squeezed vacuum state (solid line)
inputs as function of storage time $t_s$. $n$ denotes mean number of photons.
}
\end{figure}


\begin{thebibliography}{99}



\bibitem{cryptography} C. H. Bennett and G. Brassard, Proc. of IEEE Int.
Conf. on Comp. Systems and Signal Processing, Banglore India (IEEE, New York
1984); A. K. Ekert, Phys. Rev. Lett. {\bf 67}, 661 (1991).

\bibitem{teleportation} C. H. Bennett {\it et al.}, Phys. Rev. Lett. {\bf 70}, 
1895 (1990); B. Bouwmeester {\it et al.}, Nature {\bf 390}, 575 (1997);
D. Boschi et al.,  Phys. Rev. Lett. {\bf 80}, 1121 (1998); 
A. Furusawa et al. Science {\bf 282}, 706 (1998).

\bibitem{Q-comp}  A. Ekert, in {\it Proc. ICAP}, ed. by D. Wineland et al.
(AIP Press, New York, 1995), p.450.; 
 R. P. Feynman, Int. J. Theor. Phys. {\bf 21}, 467
(1982); D. Deutsch, Proc. R. Soc. London A {\bf 425}, 73 (1989);
 J. I. Cirac and P. Zoller, Phys. Rev. Lett. {\bf 74}, 4091 (1995).


\bibitem{Gisin_exp} W. Tittel {\it et al.}, Phys. Rev. Lett.
{\bf 81}, 3563 (1998).

\bibitem{ol98} M. D. Lukin et al., Opt. Lett. {\bf 23}, 295 (1998).


\bibitem{dark} For reviews on dark states
and EIT see: E. Arimondo, Progr. in Optics 
{\bf 35}, 259 (1996); S. E. Harris, Physics Today {\bf 50}, 36 (1997).

\bibitem{spe} D.J. Wineland, {\it et al.}, Phys.Rev.A {\bf 46}, R6797 (1992); 
{\it ibid} {\bf 50}, 67 (1994).

\bibitem{group}
(a) L.\ V.\ Hau, S.\ E.\ Harris, Z.\ Dutton, and C.\ H.\ Behroozi,
    Nature \textbf{397}, 594 (1999); (b) M. Kash {\it et al.}
Phys. Rev. Lett. {\bf 82}, 5229 (1999); (c) D.Budker {\it et al.},
{ibid} {\bf 83}, (1999).


\bibitem{QED}  A. S. Parkins {\it et al.}, Phys. Rev. Lett. {\bf 71}, 
3095 (1993);  T. Pellizzari {\it et al.}, {ibid}  {\bf 75}, 3788 (1995).

\bibitem{Polzik} A.~Kuzmich, K.~M\o lmer, and E.~S.~Polzik, 
Phys. Rev. Lett. {\bf 79}, 4782 (1997); 
J.~Hald, {\it et al.}, {ibid} 1319 (1999).

\bibitem{Marlan_fest} For a detailed discussion of {\it 
single-photon} trapping 
see: M. Fleischhauer, S. F. Yelin, and M. D. Lukin, Optics Comm., in press.



\bibitem{Siegman} A. Siegmann, {\it Lasers}, (University Science Books,
Mill Valley CA, 1986).


\bibitem{adiab} M. Fleischhauer and A. S. Manka, Phys. Rev. A 
{\bf 54}, 794 (1996);
K. Bergmann, H. Theuer, and B. W. Shore, Rev. Mod. Phys. {\bf 70}, 1003 (1998).



\bibitem{NLO} 
  A.\ Imamo\u{g}lu, {\it et al} 
    \prl \textbf{79}, 1467 (1997); S.\ E.\ Harris, Y.\ Yamamoto,
    {\it ibid} \textbf{81}, 3611 (1998);
 M.\ D.\ Lukin {\it et al},
    {\it ibid}  \textbf{82}, 1847 (1999);
S.\ E.\ Harris and L.\ V.\ Hau, {\it ibid} \textbf{82},
4611 (1999); A.~S.~Zibrov, M.~D.~Lukin, and M.~O.~Scully,  
{\it ibid} \textbf{83}, 4049 (1999).

\bibitem{Parkins99} A. S. Parkins and H.J. Kimble, J. Opt. B:
Quantum Semicl. Opt. {\bf 1} (1999) 496.








\end{thebibliography}
\end{document}